\documentclass[a4paper,11pt]{llncs}
\usepackage[utf8]{inputenc}
\usepackage{graphicx}
\usepackage[margin=1.5in]{geometry}

\usepackage{xcolor}
\usepackage[caption=false]{subfig}
\usepackage{amsmath,amssymb,amsfonts,flushend,tabularx}

\newcommand{\bparagraph}[1]{\par\noindent\textbf{#1}}
\newcommand{\highlight}[1]{{#1}}
\begin{document}

\title{{\mdseries\textsc{EnCoD}}: Distinguishing Compressed and Encrypted 
       File Fragments}

\author{Fabio De Gaspari\inst{1} \and Dorjan Hitaj\inst{1}
Giulio Pagnotta\inst{1} \and Lorenzo De Carli\inst{2} \and Luigi V. Mancini\inst{1}
}

\institute{Dipartimento di Informatica, Sapienza Universit\`{a} di Roma, Italy\\ 
\email{\{degaspari, hitaj.d, pagnotta, mancini\}@di.uniroma1.it},\\ 
\and
Department of Computer Science, Worcester Polytechnic Institute\\
\email{ldecarli@wpi.edu}}

\maketitle

\begin{abstract}
  Reliable identification of encrypted file fragments is a requirement for several security applications, including ransomware detection, digital forensics, and traffic analysis. A popular approach consists of estimating high entropy as a proxy for randomness. However, many modern content types (e.g. office documents, media files, etc.) are highly compressed for storage and transmission efficiency. Compression algorithms also output high-entropy data, thus reducing the accuracy of entropy-based encryption detectors. 

  Over the years, a variety of approaches have been proposed to distinguish encrypted file fragments from high-entropy compressed fragments. However, these approaches are typically only evaluated over a few, select data types and fragment sizes, which makes a fair assessment of their practical applicability impossible. 
  This paper aims to close this gap by comparing existing statistical tests on a large, standardized dataset. Our results show that current approaches cannot reliably tell apart encryption and compression, \emph{even for large fragment sizes.} To address this issue, we design \textsc{EnCoD}, a learning-based classifier which can reliably distinguish compressed and encrypted data, starting with fragments as small as 512 bytes. We evaluate \textsc{EnCoD} against current approaches over a large dataset of different data types, showing that it outperforms current state-of-the-art for most considered fragment sizes and data types.
\end{abstract}


\section{Introduction}
\label{sec:Introduction}


Reliable detection of encrypted data fragments is an important primitive with many applications to security and digital forensics. For instance, ransomware detection algorithms use estimates of write-operations' data randomness to quickly identify evidence of malicious encryption processes~\cite{kirda_redemption,mehnaz_rwguard,continella_shieldfs:_2016,kirda_unveil}. When performing digital forensic analysis of hard drives and phones, it is oftentimes important to identify encrypted archives~\cite{conti_automated_2010}. Finally, encryption detection is widely used in network protocol analysis~\cite{de_carli_botnet_2017,dorfinger_real-time_2010}. 

A popular approach to address this problem is to estimate the Shannon entropy of the sequence of interest using the Maximum Likelihood Estimator (MLE):  $\hat{H}_{MLE}$. This approach leverages the observation that the distribution of byte values in an encrypted stream closely follows a uniform distribution; therefore, high entropy is used as a proxy for randomness. This estimator has the advantage of being simple and computationally efficient. As non-encrypted digital data is assumed to have low byte-level entropy, the estimator is expected to easily differentiate non-encrypted and encrypted content.

While this approach remains widely used (e.g.,~\cite{kirda_redemption,mehnaz_rwguard,continella_shieldfs:_2016,kirda_unveil}), a number of works have highlighted its limitations. Modern applications tend to compress data prior to both storage and transmission. Popular examples include the zip compressed file format, and HTTP compression~\cite{http-rfc} (both using the DEFLATE algorithm). As compression removes recurring patterns in data, compressed streams tend to exhibit high Shannon entropy. As a result, compressed data exhibit values of $\hat{H}_{MLE}$ that are close and oftentimes overlapping with those obtained by encryption. In principle, compressed content can be identified by using appropriate parsers. However, many security-related applications, such as ransomware detection, traffic analysis and digital forensics, generally do not have access to whole-file information, but rather work at the level of \textit{fragments} of data. In these settings, the metadata that is required by parsers is not present or is incomplete~\cite{park_data_2008}.
Given this issue, a number of works have been looking at alternative tests to distinguish between encrypted and compressed content~\cite{malhotra_detection_2007,wang_using_2011,foresti_efficient_2016,lipmaa_data_2017,hahn2018detecting,casino_hedge_2019,choudhury_empirical_2020}. While these works have the potential to be useful, there has been limited evaluation of their performance on a standardized dataset. Consequently, there is no clear understanding of how these approaches: (i) fare on a variety of compressed file formats and sizes, and (ii) compare to each other. The potential negative implications are significant: the use of ineffective techniques for identifying encrypted content can hinder the effectiveness of ransomware detectors~\cite{gaspari2019}, and significantly limit the capability of forensic tools.

Our work compares select state-of-the-art approaches on a large dataset of different data types and fragment sizes. We find that, while more useful than entropy estimates, current approaches fail to achieve consistently high accuracy. To address this, we propose \textsc{EnCoD} (\textbf{En}cription/\textbf{Co}mpression \textbf{D}istinguisher), a novel neural network-based approach. Our evaluation shows that \textsc{EnCoD} outperforms existing approaches for most considered file types, over all considered fragment sizes. \textsc{EnCoD} can classify data fragments as small as $512B$ with $86\%$ accuracy, increasing to up to $~94\%$ for purely compressed data (i.e., zip, gzip) and up to $100\%$ for compressed application data fragments (e.g., pdf, jpeg, mp3) for $8KB$ fragment sizes. 
Overall, this paper makes the following contributions:

\begin{itemize}
\item We review and categorize existing literature on the topic of distinguishing compressed and encrypted data fragments.
\item We systematically evaluate and compare state-of-the-art approaches on a large, standardized dataset including a variety of fragment formats and sizes.
\item We propose a new neural-network based approach, which outperforms current state-of-the-art tests in distinguishing encrypted from compressed content for most considered formats, over all considered fragment sizes.
\item We propose a new multi-class classifier that can label a fragment with high accuracy as encrypted data, general-purpose compressed data (zip/gzip/rar), or one of multiple application-specific compressed data (png, jpeg, pdf, mp3).
\item We thoroughly discuss the implications of our findings, in terms of the applicability of the evaluated approaches.
\end{itemize}







\section{Background}
\label{sec:Background}

Determining the format of a particular data object (e.g. a file in permanent storage, or an HTTP object) is an extremely common operation. Under normal circumstances, it can be accomplished by looking at content metadata or by parsing the object. Things get more complicated, however, when no metadata is available and the data object is corrupted or partly missing. In this paper, we focus on detection of \textit{encrypted content}, in particular on distinguishing between encrypted and compressed data fragments.  We begin by examining relevant applications of encryption detection primitives.

\subsection{Ransomware Detection}

\textit{Ransomware} encrypts user files with the aim of making them unusable for the user. It then presents a prompt asking the user to pay a ransom in order to receive the decryption key. Ransomware attacks can cause significant financial damage to organizations~\cite{ransomware_atlanta:2018,ransomware_uk:2017,ransomware_trends_nyt}.

Mitigating a ransomware infection requires rapid detection and termination of all ransomware processes. 
A number of approaches based on \textit{behavioral process analysis} have been proposed for this purpose~\cite{kirda_redemption,mehnaz_rwguard,continella_shieldfs:_2016,kirda_unveil}. These approaches typically use a classifier, based on various features of process execution, to distinguish benign and ransomware processes. Virtually all proposed behavioral detectors use entropy of file write operations as a key feature, based on the insight that frequently writing encrypted content is a characteristic behavioral fingerprint of ransomware. Entropy is typically estimated using $\hat{H}_{MLE}$. In several approaches entropy is estimated on the content of individual file writes~\cite{kirda_redemption,kirda_unveil,continella_shieldfs:_2016}, therefore the estimation procedure has only access to partial file fragments.

\subsection{Forensics}

Digital forensics oftentimes involves analysis of phone~\cite{walls_forensic_2011} or PC~\cite{park_data_2008} storage that has been corrupted, or uses an unknown format. Therefore, forensic techniques attempt to recover data of interest (contacts, pictures, etc.) by searching for blocks with recognizable structure. These techniques typically only have access to data fragments, rather than whole files.

Encrypted and compressed data represent a corner case, as they exhibit a complete lack of structure. Still, detecting such content may be important in data recovery operations (e.g., if sensitive data is known to have been encrypted). Distinguishing between compressed and encrypted blocks is notoriously difficult, and some forensic approaches label data as \textit{``compressed or encrypted''}, without attempting to pinpoint which one of the two it is~\cite{conti_automated_2010}. 

\subsection{Network Traffic Analysis}

Network traffic analysis examines flows in/out of a network to identify security issues. Regulations (e.g. HIPAA in the U.S.) and best practices expect sensitive data to be encrypted in transit; therefore, entropy-based analyzers have been proposed to ensure that all traffic leaving a monitored network is encrypted~\cite{dorfinger_real-time_2010}. Another application is reverse-engineering of network protocols used by malware. It has been observed~\cite{de_carli_botnet_2017} that malware protocols may mix encrypted and unencrypted content within the same message. Encryption detection primitives can be applied to break messages into encrypted and unencrypted fields.

In both cases above, encryption detectors have partial visibility on the data stream and can only access fragments of data (e.g., an encrypted stream broken into individual packets), rather than whole data objects.

\subsection{Challenges}
\label{sec:Challenges}

In the three domains above, the use of Shannon entropy has been proposed in order to identify encrypted content. Entropy is used to measure the \textit{information content} of a byte sequence; highly structured data exhibit low entropy, while unstructured data---such as a randomly distributed sequence---have high entropy. Therefore, an entropy estimate can be used as a proxy for how close a sequence of bytes is to being randomly distributed. Most encryption algorithms output ciphertexts whose byte-value distributions tend to follow a uniform distribution. As a result, an encrypted bytestream will almost invariably exhibit high entropy.

One of the most common approaches to entropy estimation is the Maximum Likelihood Estimator $\hat{H}_{MLE} = -\sum_{i=0}^{255}{f_ilog_2(f_i)}$, where $f_i$ is the frequency of byte value $i$ in the sequence.
The entropy range is $[0-8]$. The frequency $f_i$ of byte value $i$, which is measurable, is used in place of the probability $P(i)$ of that value occurring, which is unknown. 
This approach is commonly used in some of the applications above (e.g.,~\cite{kirda_redemption,dorfinger_real-time_2010}), due to its simplicity and efficiency.

This reasoning assumes that, while encrypted data has high entropy, non-encrypted data does not. This appears reasonable, as most relevant data types (e.g., text, images, audio) are information-rich and highly structured. However, this assumption does not hold true in modern computing. Modern CPUs can efficiently decompress data for processing, and compress it back for storage or transmission; this is oftentimes performed in real time and transparent to the user. As a result, most formats tend to apply compression~\cite{noauthor_docx_2017,wallace1992jpeg}. Informally, a good compression algorithm works by identifying and removing recognizable structures from the data stream; as a result, compressed data tend to exhibit high entropy. In practice, this fact compromises the ability of entropy-based detectors to distinguish encrypted and non-encrypted, compressed content.

\begin{figure}[t]
  \centering
  \includegraphics[width=0.75\textwidth]{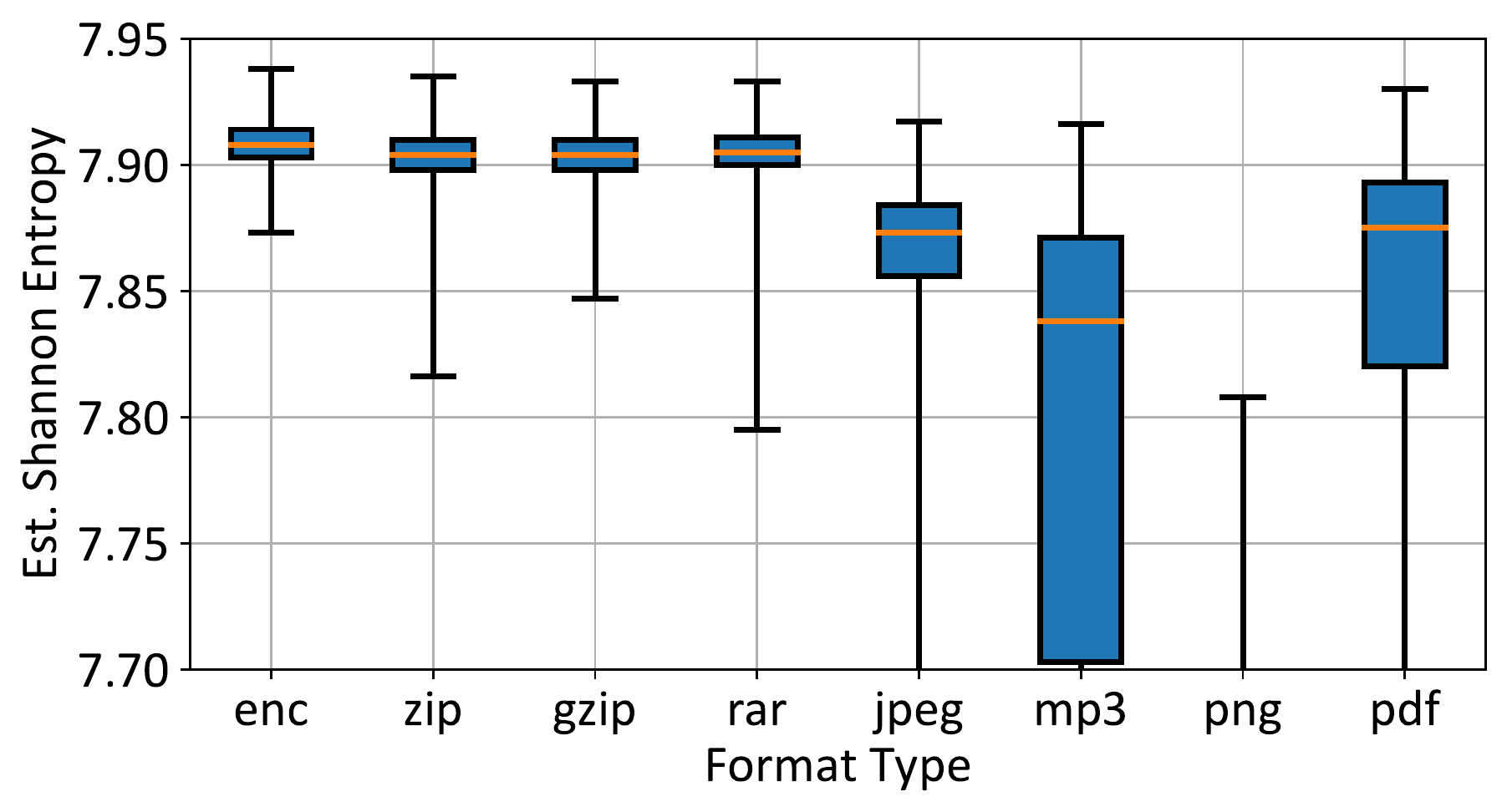}
  \vspace{-0.1in}
  \caption{Entropy ranges for common formats (2048B blocks)}
  \label{fig:entro}
  \vspace{-0.1in}
\end{figure}

  \bparagraph{Entropy estimates for common data formats.} In order to substantiate the claim above, we computed entropy estimates using a dataset consisting of \textbf{10,000} file fragments. The dataset covers various popular file formats and AES-256-encrypted data. We considered multiple fragment sizes, from 512B to 8KB (details in Section~\ref{sec:Classifier}). Figure~\ref{fig:entro} summarizes the distribution of estimated entropy values for eight different formats with block size 2048 (some ranges truncated for clarity). Results for other block sizes were qualitatively similar; full results are tabulated in Appendix~\ref{sec:EntropyAnalysisResults}. As illustrated in Figure~\ref{fig:entro}, both general-purpose (zip, rar) and domain-specific (jpeg, mp3) compression algorithms result in data which exhibits entropy whose ranges are overlapping with that of encrypted content (enc). The only format that can be unambiguously distinguished is png. Even so, png still overlaps with various other formats. Interestingly, utilities that create and modify data in zip, gzip and png format internally all use the DEFLATE algorithm for compression: the differences in entropy are likely due to differences in file structure and algorithm implementation.

Due to the limits of entropy estimation, the attention of the community has been increasingly focusing on alternative measures that can more precisely estimate whether data follow a random distribution. However, no comprehensive review of such approaches exist. In the next section, we review state-of-art approaches, while we evaluate and compare them in Section~\ref{sec:Evaluation}.

\section{Review of Existing Techniques}
\label{sec:Review}

This section reviews three state-of-the-art approaches to distinguishing encrypted and compressed content: the NIST suite, $\chi^2$ and HEDGE~\cite{casino_hedge_2019}. Strictly speaking, these approaches test the \textit{randomness} of a string of bytes, and make no attempt to determine its type. However, due to their high precision they can be used to distinguish true pseudorandom (encrypted) sequences and compressed ones which, while approximating a randomly generated stream, maintain structure. 

The NIST suite and $\chi^2$ are standard statistical tests for identifying randomly-distributed data. HEDGE is a recently proposed statistical approach which shows promising results. HEDGE is a combination of a subset of the NIST tests and two forms of $\chi^2$ tests. Note that, despite the inclusion of HEDGE, we decided to also report separate results for NIST and $\chi^2$ due to the fact that those are designed to be, and oftentimes are, used as standalone tests.

\subsection{NIST SP800-22}
\label{sec:ReviewNist}

The NIST SP800-22 specification~\cite{rukhin_statistical_2010} describes a suite of tests whose intended use is to evaluate the quality of random number generators. The suite consists of 15 distinct tests,  which analyze various structural aspects of a byte sequence. These tests are commonly employed as a benchmark for distinguishing compressed and encrypted content (e.g.,~\cite{casino_hedge_2019,choudhury_empirical_2020}). 
Each test analyzes a particular property of the sequence, and subsequently applies a test-specific decision rule to determine whether the result of the analysis suggests randomness or not. 

When using the NIST suite for discriminating random and non-random sequences, an important question concerns aggregation of the results of individual tests. Analysis of the tests~\cite{rukhin_statistical_2010} suggests that they are largely independent. Given this observation, and the intrinsic complexity of \textit{a priori} defining a ranking between the tests, we use a \textit{majority voting} approach. In other words, we consider a fragment to be random (and therefore encrypted) when the majority of tests considers it so. Since some of the tests require a block length much bigger than the ones we use for our smaller fragment sizes, we did not consider in the voting the tests that cannot be executed.

\subsection{$\chi^2$ Test}
\label{sec:ReviewChi2}

The $\chi^2$ test is a simple statistical test to measure goodness of fit. It has been widely applied to distinguish compressed and encrypted content~\cite{malhotra_detection_2007,lipmaa_data_2017,casino_hedge_2019}. Given a set of samples, it measures how well the distribution of such samples follows a given distribution. 
Mathematically, the test is defined as:

\vspace{0.1in}
$\chi^2 = \sum\limits_{i=0}^{255}\frac{(N_i - E_i)^2}{E_i}$
\vspace{0.1in}

\noindent where $N_i$ is the actual number of samples assuming value $i$, and $E_i$ is the expected number of samples assuming value $i$ according to the known distribution of interest. Since the distribution being evaluated for goodness of fit is the discrete uniform distribution, $\forall i E_i = L/256$, where $L$ is the particular fragment length being considered. The results of the test can be interpreted using either a fixed threshold, or a confidence interval~\cite{casino_hedge_2019}.


\subsection{HEDGE}
\label{sec:hedgepaper}

HEDGE~\cite{casino_hedge_2019} simultaneously incorporates three methods to distinguish between compressed and encrypted fragments: $\chi^2$ test with absolute value, $\chi^2$ with confidence interval and a subset of NIST SP800-22 test suite. Out of the NIST SP800-22 test suite HEDGE incorporates 3 tests: \textit{frequency within block test}, \textit{cumulative sums test}, and \textit{approximate entropy test}. These tests were selected due to (i) their ability to operate on short byte sequences, and (ii) their reliable performance on a large and representative dataset. In the HEDGE detector the threshold of the number of the above-mentioned NIST SP800-22 tests failed is set to 0.
For the $\chi^2$ with absolute value test, the thresholds are pre-computed for each of the considered packet sizes, by considering the average and its standard deviation.
For $\chi^2$ with confidence interval, the $\chi\%$ interval is $(\chi\% > 99\% || \chi\% < 1\%)$.
For classifying the content of a packet, HEDGE applies the three randomness tests to the input data. Data is considered random only if it passes all tests.




\section{{\mdseries\textsc{EnCoD}}: A Learning-based Approach}
\label{sec:Classifier}

%
Past work and our own evaluation suggest that tests based on byte-value distribution, such as $\chi^2$, can distinguish some encrypted and compressed content, but have accuracy issues (ref. Section~\ref{sec:Evaluation}). Such tests, in a sense, ``collapse'' the entire distribution to a single scalar value, losing information concerning the shape of the distribution. It is therefore natural to ask if Deep Neural Networks (DNNs) can improve such results. DNNs can consider the entire discrete distribution (modeled as a feature vector), and can learn to recognize complex distributions~\cite{Lee0MRA17}.
%
%
 In order to evaluate the potential of DNNs we designed \textsc{EnCoD} (\textbf{En}cryption/\textbf{Co}mpression \textbf{D}istinguisher), a set of two distinct neural network-based approaches for distinguishing encryption and compression. 



\vspace{-0.1in}
\subsection{Model Architecture \#1: Binary Classifiers}
\label{sec:mabc}
\vspace{-0.05in}

Our first model is a binary classifier trained to distinguish a single specific compressed format from encrypted content. It may be used in cases where only one compressed format is known to exist in the dataset (e.g., detecting writes of encrypted data performed by a potential ransomware on image files vs legitimate writes of JPEG-compressed data).
We explored several alternative architectures for this application, and we found that the structure depicted in Figure~\ref{fig:binary_model} provides the best performance. The binary-classifier architecture consists of 4 fully-connected layers with dimensions as shown in the figure. We initialize the model weights using Glorot uniform~\cite{Glorot2010UnderstandingTD}. The activation function is ReLU for the first 3 layers, followed by a softmax on the output layer. We used a batch size of $64$ for training our model. Each hyperparameter has been chosen using grid search. We used the same procedure also for the model described in Section~\ref{sec:contenttypedet}.

\vspace{-0.1in}
\subsection{Model Architecture \#2: Content-Type Detector}
\label{sec:contenttypedet}
\vspace{-0.05in}

\begin{figure*}[t]
  \centering
  \subfloat[Binary Classifier Architecture\label{fig:binary_model}] {
     \includegraphics[width=.8\linewidth]{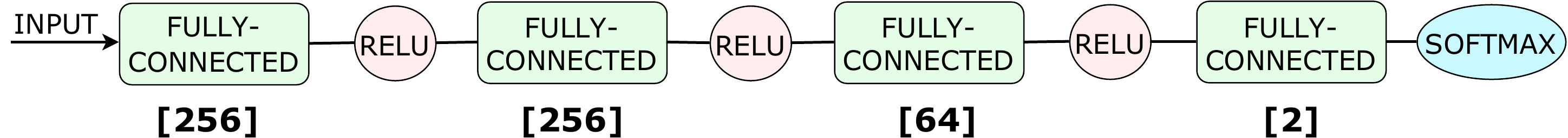}
  }
  \hfill
  \subfloat[Multi-Class Classifier Architecture\label{fig:multi_model}]{
     \includegraphics[width=1\linewidth]{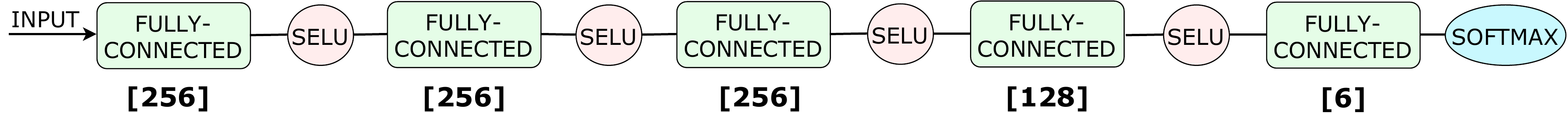}
  }
  \caption{Neural Network Architectures}
  \label{fig:model_arch}
  \vspace{-0.15in}
 \end{figure*}

In many applications, a classifier may encounter more than one type of compressed data. Furthermore, it may be important to determine the specific type being encountered. To support these use cases, we design a \textit{content-type detector}: a multi-class classifier that can determine whether a given fragment is encrypted, or belongs to one of multiple known compressed formats. We explored several designs for the neural network, converging to the model depicted in Figure~\ref{fig:multi_model}. Its architecture consists of 5 fully-connected layers with dimensions as shown in the figure. We initialize model weights using LeCun normal~\cite{LeCun1998EfficientB}. Differently from the binary models, this multi-class classifier seemed prone to the dying neuron problem associated with the ReLU activation function~\cite{8260635}. We therefore opted for the SelU activation function~\cite{DBLP:journals/corr/KlambauerUMH17} for the first 4 layers, followed by a softmax on the output layer. We used a batch size of $64$ instances for training.




\subsection{Fragment Dataset}
\label{sec:file_dataset}

\highlight{We built a dataset of \textbf{200M encrypted and compressed fragments}. For the compressed data, we selected a set of formats covering common, popular content types. To generate the encrypted data fragments, we used the AES cipher in CBC mode implemented by the PyCryptodome library~\cite{pycrypto}. We chose AES because it is the most widely used and well known symmetric cipher, representative of modern ciphers which result in byte streams consistently close to random.

In constructing the dataset, we focused on ensuring a diversity of compressed formats, rather than compression algorithms. While algorithms such as DEFLATE are used in multiple compression formats, they are generally used with different parameters and/or embed compressed data in different ways within the compressed archive. Consequently, compressed archives created with different formats tend to differ considerably from each other even when using the same underlying compression algorithm.
This observation is empirically confirmed by our evaluation in Section~\ref{sec:Evaluation}. Finally, our dataset does not include data which is both compressed and encrypted, and we ensured such data is not present in the dataset. The dataset is comprised of the following data types:}

\begin{enumerate}
    \item \textbf{AES encrypted data (enc)}. \highlight{We used the AES implementation provided by the Cryptodome Python library. AES was configured to use CBC mode with 256-bit keys, with a random IV generated before encrypting each file}.
\item \textbf{DEFLATE- and rar-compressed data (zip/gzip and rar)}: both DEFLATE and rar are de-facto standards for generic file compression. DEFLATE is also widely used for documents (such as in the MS OFFICE file formats), and network applications (e.g., HTTP header compression).
\item \textbf{png and jpeg images:} png is used for lossless image compression; it internally uses DEFLATE, but png files present a structure that is different different from that of zip files. jpeg uses DCT-based lossy compression.
\item \textbf{mp3 audio files:} MP3 compressors use a psychoacoustic model to remove inaudible frequencies from audio data, and compress the resulting data using a lossy algorithm based on the modified-DCT transform.
\item \textbf{pdf documents:} PDF is an office format used for document exchange and form filling. Internally, PDF files consist of a tree of objects that can be compressed using a variety of techniques. In practice, most PDF documents contain a large amount of compressed content, such as embedded images.
\end{enumerate}

\vspace{-0.15in}
\subsubsection{Fragment generation process.}

We generate fragments from a dataset of files:

\begin{itemize}
\item \textbf{zip/gzip/rar/enc:} we used various textual documents obtained from a 2020 English Wikipedia dump~\cite{wiki_dump}. We created four copies of each file, each of which was either compressed using one of zip, gzip, rar utilities (with default parameters), or encrypted using AES-256.
\item \textbf{png:} we crawled 5000 png images from the web and various repositories~\cite{png_images}.
\item \textbf{jpeg: } we downloaded 10,000 images from the Open Images Dataset v5~\cite{jpeg_images}.
\item \textbf{mp3:} we used the FMA dataset~\cite{mp3_files}, which contains 8000 mp3 files.
\item \textbf{pdf:} we crawled 1,000 randomly-selected papers from arXiv~\cite{arxiv_2019}.
\end{itemize}

We split each file into fragments of 512B, 1KB, 2KB, 4KB, and 8KB. We then selected 5M fragments for each fragment size/data type combination (this ensures that the dataset remains balanced).

\subsection{Dataset Analysis Methodology}


\paragraph{Statistical tests (NIST, $\chi^2$, HEDGE):} For each fragment size, we randomly selected 10,000 compressed fragments (evenly distributed across the different compressed data types) and 10,000 encrypted fragments. We then executed the tests directly on these fragments.

\paragraph{\textsc{EnCoD}/Binary Classifiers:}

We separately trained and evaluated classifiers for each fragment size. The features that are fed to our models for training/classification are derived from the histograms of the byte values for the observed fragment size. Each feature is the value of the probability density function at a given bin, normalized such that the integral over the range is 1.

We trained the binary classifiers by randomly selecting 3M vectors from the encrypted class and 3M vectors from the data type that we aim to distinguish. We partitioned this dataset into 85\% training, 5\% development and 10\% test. Before fitting the data to the model for training, we applied a MinMax scaler to scale the dataset from the range $[0,1]$ to the range $[0,2]$ (range selected via grid search). Scaling helps the ML model to more easily capture minute differences in the inputs, allowing to better distinguish among the classes and converge faster. 

\paragraph{\textsc{EnCoD}/Content-Type Detector:}

To train the content-type detectors, for each fragment size we randomly sampled 6M feature vectors consisting of a mix of the considered file types. This dataset was partitioned into training, development and test sets in the same ratios used for the binary classifiers. We also scaled the dataset using the MinMax scaler with the same parameters used above.

\section{Evaluation }
\label{sec:Evaluation}

In this section, we comprehensively evaluate existing approaches discussed in Section~\ref{sec:Review}, in addition to \textsc{EnCoD}, our novel neural network-based approach (see Section~\ref{sec:Classifier}). We frame the evaluation in terms of the following comparisons:

\begin{enumerate}
\item \textbf{Binary classification: all formats.} In Section~\ref{sec:bcaf}, we consider the ability of different detectors to discriminate encrypted and compressed data, regardless of the specific compressed format. Results show that our classifier heavily outperforms NIST, $\chi^2$-test and HEDGE for all fragment sizes.
\item \textbf{Binary classification by format.} In Section~\ref{sec:bcbf}, we break down the performance of $\chi^2$, NIST and HEDGE by compressed format. We also report the performance of our per-format binary classifiers (see Section~\ref{sec:mabc}). The latter perform comparably or better than other tests on all formats but one.
\item \textbf{Format fingerprinting.} In Section~\ref{sec:ff}, we evaluate the accuracy of our multi-class classifier in labeling unknown fragments as the correct compressed format (or as encrypted). Results show that our classifier is able to distinguish the file type with an overall accuracy of 90\% for the 2048 byte chunk size. It also achieves high precision, especially on png, jpeg, mp3.
\end{enumerate}

\subsection{Implementation}

We implemented the classifier described in Section~\ref{sec:Classifier} using the Keras Library~\cite{chollet2015keras} for machine learning.  For the NIST tests, we used the official implementation~\cite{computer_security_division_nist_2016}. In order to aggregate the NIST tests results, we use the majority voting approach described in Section~\ref{sec:ReviewNist}. In order to label fragments as compressed or encrypted based on $\chi^2$ results, we used the thresholds suggested in the HEDGE paper~\cite{casino_hedge_2019}, as the analysis in HEDGE is specifically aimed at producing a dataset-independent threshold for general use. We implemented HEDGE according to the published description~\cite{casino_hedge_2019}. Finally, all experiments were conducted using the dataset described in Section~\ref{sec:Classifier}.
\begin{figure}[t]
  \centering
  \includegraphics[width=0.7\textwidth]{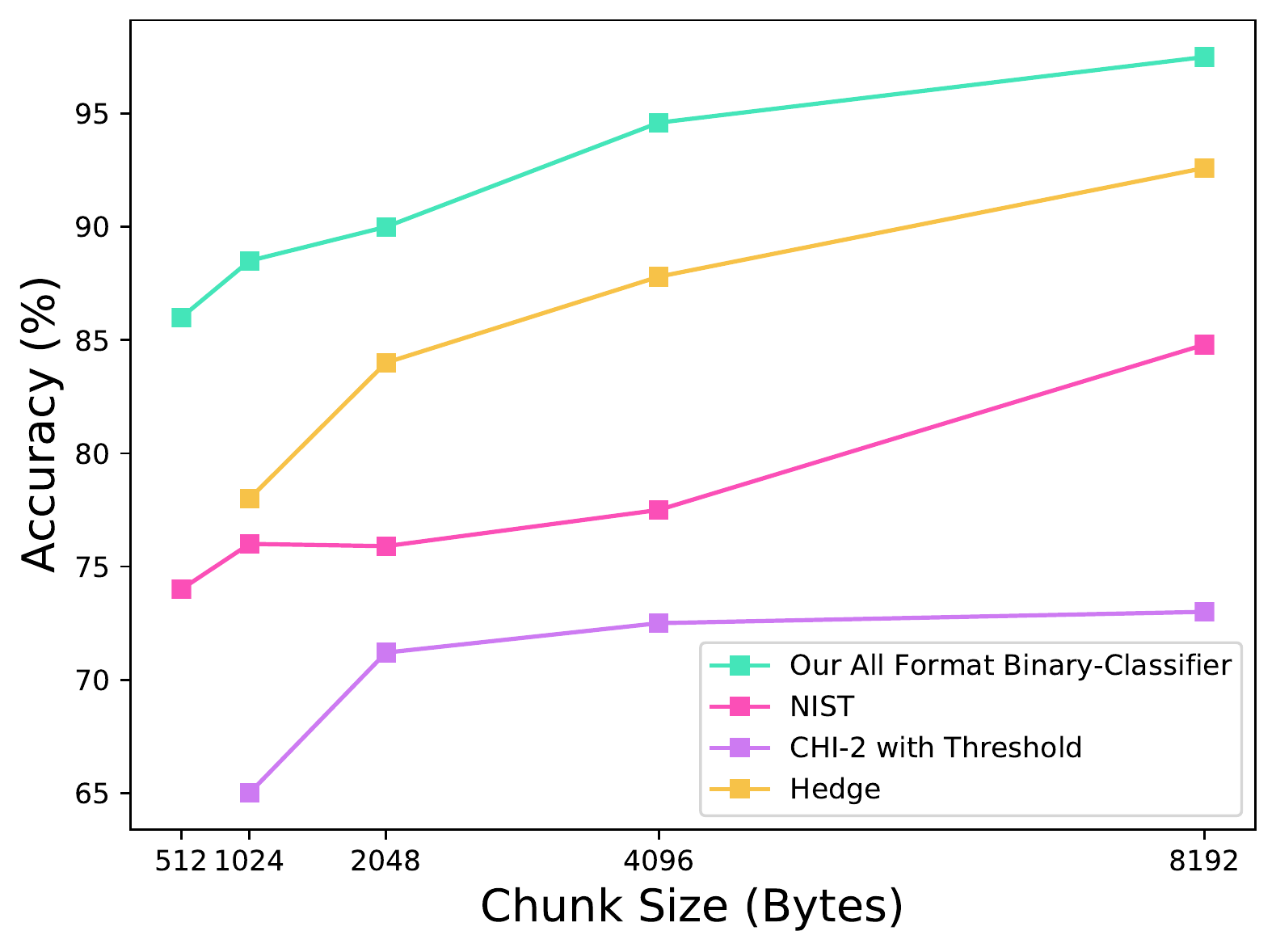}
  \caption{Performance comparison (binary classification: all formats) }
  \label{fig:bcaf}
\end{figure}
\subsection{Binary Classification: All Formats}
\label{sec:bcaf}

In the first part of our evaluation, we consider the binary classification problem of determining whether a given high-entropy data fragment is compressed or encrypted. Given a fragment, the $\chi^2$ test, HEDGE, and the NIST test suite return whether the fragment's content appears random or not. Therefore, a binary classifier can be derived simply by labeling random content as encrypted. Our binary classifier used for this evaluation is based on our multi-class classifier. The multi-class classifier labels each fragment either as encrypted, or as one of the seven supported compressed formats. Since in this experiment we are only interested in distinguishing encryption and compression, regardless of the type, we combine all compressed type labels into one (we look at content fingerprinting accuracy in Section~\ref{sec:ff}). Effectively, we consider classification in two labels: (1) a macro-label ``compressed'', which is comprised of the labels $\{zip, rar, gzip, png, jpeg, mp3, pdf\}$ and (2) the label ``encrypted''.

The results of this evaluation are depicted in Figure~\ref{fig:bcaf}. All classifiers tend to improve as fragment size increases; we discuss this phenomenon in Section~\ref{sec:Discussion}. Our neural network-based approach heavily outperforms all the other approaches on all block sizes. The $\chi^2$ accuracy remains consistently low across the range of block sizes. Results suggest that this test has an intrinsic difficulty in discriminating non-random content which closely approaches a uniform random distribution.

\begin{figure}[t]
  \centering
  \includegraphics[width=0.7\textwidth]{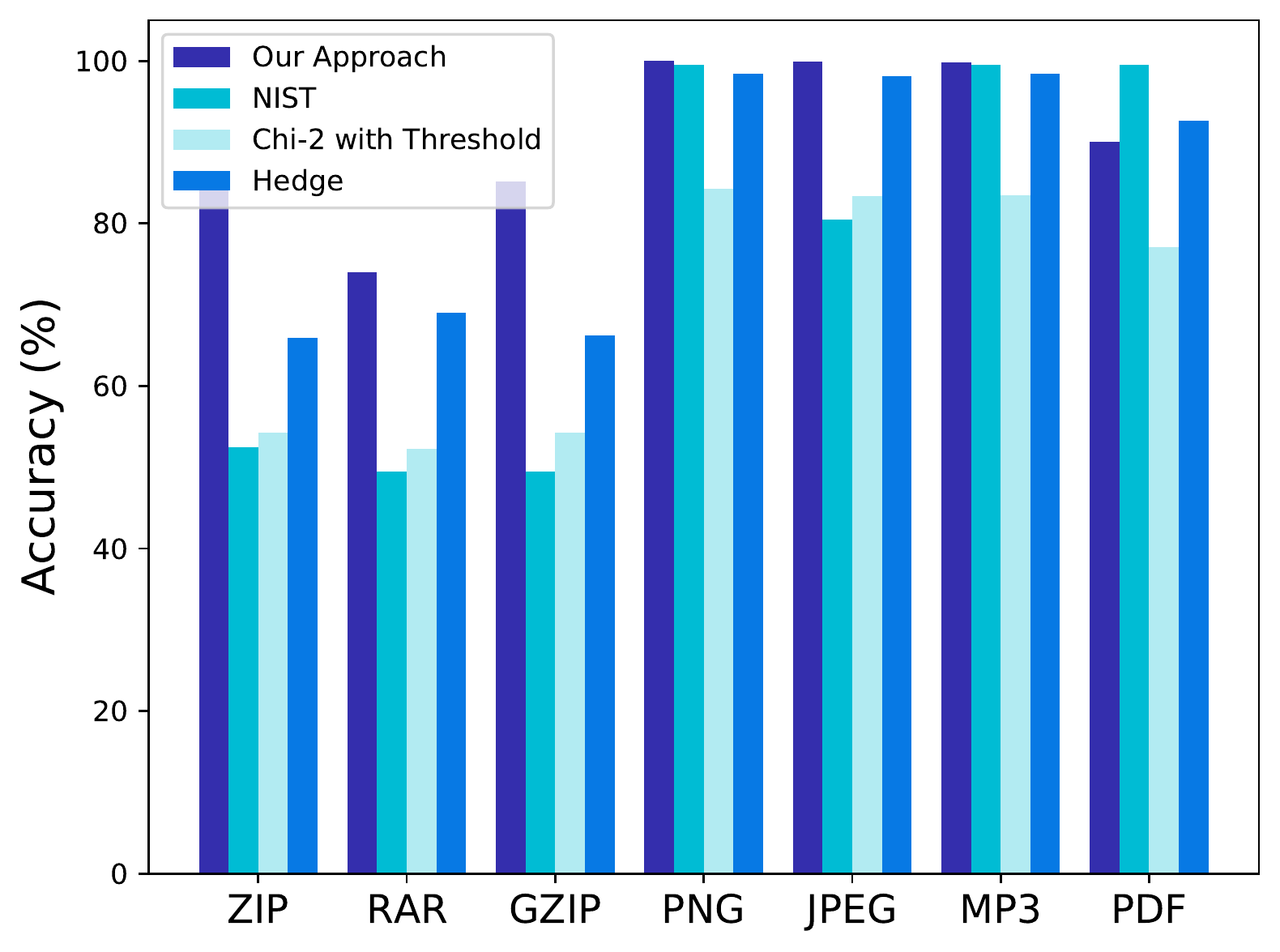}
  \caption{Performance comparison between our binary-classifier approach, NIST, $\chi^2$ with Threshold and HEDGE (2048B blocks)}
  \label{fig:bcbf}
\end{figure}

\begin{figure}[t]
  \centering
  \includegraphics[width=0.7\textwidth]{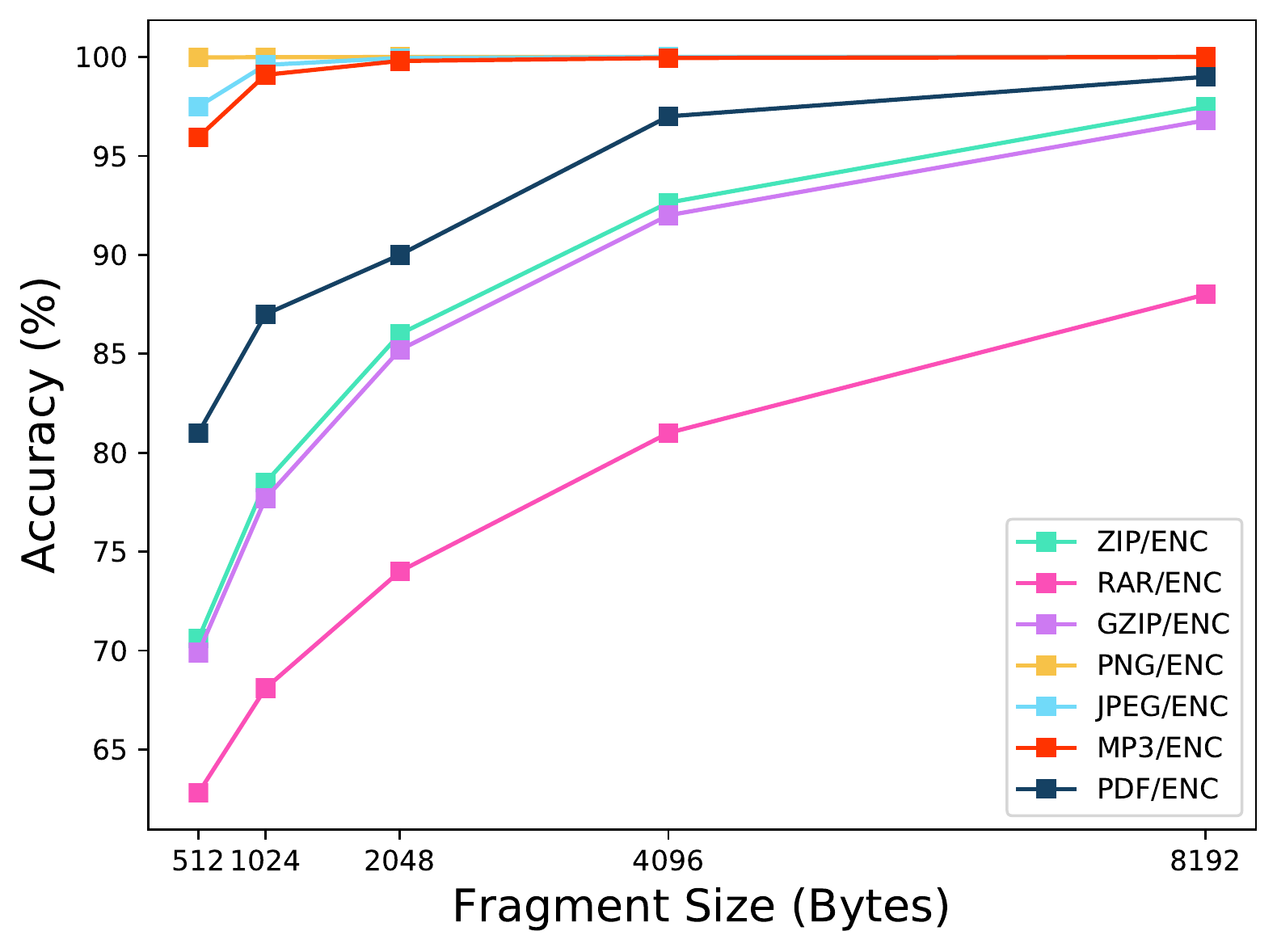}
  \caption{Performance of our binary classifiers (all block sizes)}
  \label{fig:bcbf-nn}
\end{figure}

\subsection{Binary classification by format}
\label{sec:bcbf}

In the second experiment, we consider the question of whether some compressed formats are harder than others to distinguish from encrypted content. Such phenomenon may arise due to (i) differences in effectiveness between compression algorithms in removing redundancy (and thus structure) from the uncompressed data; and (ii) presence (or absence) of metadata, or other structured information interleaved with compressed data.

In order to answer this question, we break down results for the $\chi^2$-test, NIST suite, and HEDGE test by format. We do not evaluate our multi-class classifier in this experiment. This is due to the fact that this classifier can generate two different types of classification errors for compressed formats: (1) mislabeling a compressed fragment as an encrypted one; and (2) mislabeling a compressed fragment of a given type for a compressed fragment of a different type. As $\chi^2$-test, NIST and HEDGE can only generate errors of the former type, a direct comparison is not possible.
%
Instead, in this experiment we evaluate multiple binary neural network-based classifiers (see Section~\ref{sec:mabc}). With this approach, we train one binary classifier per compressed format. Each classifier is trained to distinguish content in that format from encrypted content. It is important to note that, while each of these classifiers is trained specifically on one format, the other tests ($\chi^2$, NIST and HEDGE) work the same regardless of the format. Despite this limitation, we believe this to be an informative analysis of the potential of learning-based approaches.

Figure~\ref{fig:bcbf} shows the comparison between the three approaches on 2048-byte blocks. Overall, neural network-based classifiers tend to fare better than the other tests, particularly on challenging formats such as zip/gzip and rar. PDF is the only format on which the NIST and HEDGE tests outperform the neural network classifier. Interestingly, the $\chi^2$ fares slightly better than NIST on most formats, but its accuracy is significantly worse on formats that are typically easy to distinguish, such as PNG. We believe this to be due to the fact that the NIST tests look at a richer set of properties beyond byte value distribution, such as a presence of runs and repeated sequences. HEDGE test outperform $\chi^2$ on all file types, while outperforming NIST on most formats, beside PDF, and have similar performance on PNG and MP3 formats. Finally, Figure~\ref{fig:bcbf-nn} presents the performance of all our binary classifiers across the range of fragment sizes. These results again show that accuracy increases significantly as block size increases. For 8KB-blocks, accuracy is above 85\% for all types.

\subsection{Format Fingerprinting}
\label{sec:ff}
\begin{figure}[t]
  \centering
  \includegraphics[width=0.85\textwidth]{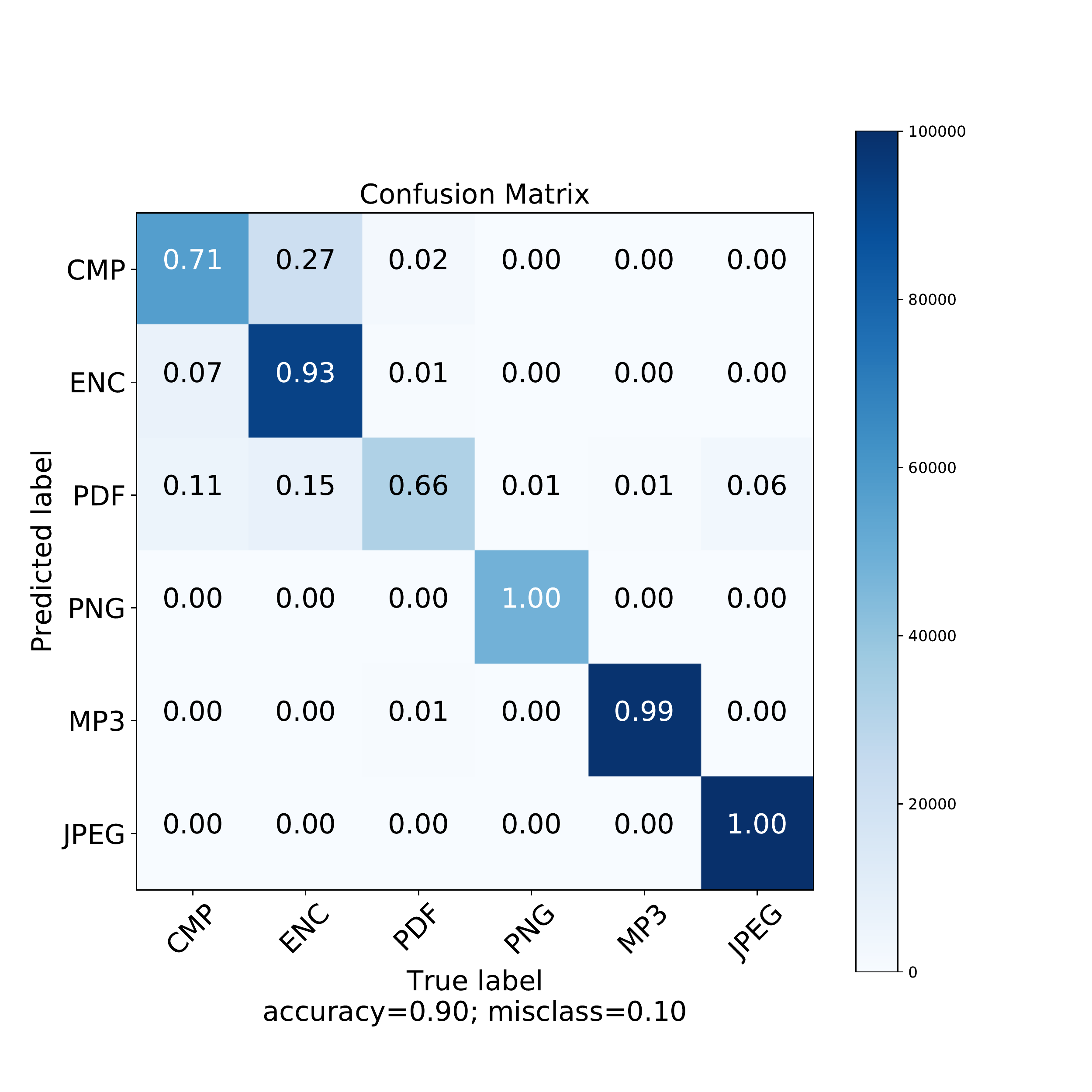}
  \vspace{-2em}
  \caption{Confusion matrix for the content-type classifier}
  \label{fig:conf_multi_class}
\end{figure}
Our multiclass classifier has the ability to (1) distinguish encrypted and compressed data, and (2) pinpoint the specific format compressed data belong to. This is a significant improvement over the functionality of existing tests, that can only distinguish encryption and compression. 
In this section, we evaluate the effectiveness of our multi-class classifier in fingerprinting the correct type of compressed content.
Figure~\ref{fig:conf_multi_class} shows the confusion matrix for the multi-class classifier. Results indicate that our classifier is able to pinpoint the file type with consistently high precision for most formats, especially png, mp3, and jpeg. It performs fairly well on the other considered compressed formats such as cmp (which contains a mixture of zip, rar, and gzip feature vectors) but with a slightly higher rate of misclassified instances between enc and cmp. This can be explained by the fact that their distributions are very close, and intrinsically hard to distinguish.

\subsection{Overhead}
\label{sec:eval_overhead}
\highlight{In the final part of our evaluation, we analyze the practical applicability of the three approaches, comparing their runtime in order to understand if they can be deployed in time-critical applications. For this test, we used a small dataset comprised of 1000 randomly-selected compressed or encrypted samples. We ran all three approaches (NIST, HEDGE and our binary ML model) on each sample, taking  individual runtime and repeating the experiment 1000 times. Table~\ref{tab:overhead} presents the results of our evaluation. As we can see, while both mean and median runtime for NIST tests are faster then HEDGE, our proposed binary classifier is considerably faster than both. Both mean and median runtime for the ML model are three orders of magnitude faster than both NIST and HEDGE, making it easily applicable to scenarios that require fast classification results such as ransomware detection. It is worth noting that the evaluation of our ML model was carried out by measuring the time required to predict a single sample, rather than a batch of samples. However, our model can easily classify multiple samples in parallel by exploiting the heavy parallelism of GPUs, further decreasing the runtime required per individual sample.}

\begin{table}[t]
\small
\centering
\begin{tabular}{|l|c|c|c|}  \hline
 \textbf{Approach}      & \textbf{Mean} & \textbf{Median}   & \textbf{Std.dev}  \\\hline
    NIST                &  0.1          &  0.1              & 0.004         \\\hline
    HEDGE               &  0.44         &  0.43             &   0.008       \\\hline
    Binary Classifier   &  0.00046      &  0.00044          &  0.00012        \\\hline
\end{tabular}
\vspace{0.1em}
\caption{Time required by each approach to classify one sample, in seconds.}
\label{tab:overhead}
\vspace{-0.2in}
\end{table}

    
    





\section{Discussion of Findings}
\label{sec:Discussion}

Results shown in Section~\ref{sec:Evaluation} highlight the difficulty of discriminating compressed and encrypted fragments. State-of-the-art statistical tests tend to fare better than entropy measures (ref. Section~\ref{sec:Background}), but their performance varies significantly depending on the specifics of the compressed format and fragment size. Moreover, such approaches can only determine whether a given fragment is encrypted with a certain confidence, but cannot distinguish between different compressed formats. \textsc{EnCoD}, the learning-based approach introduced in Section~\ref{sec:Classifier}, tackles both these limitations. Both per-format and multi-class classifiers outperform existing tests on all considered file types/block sizes. Moreover, our multi-class classifier can be used to determine the format of a given unknown fragment, even in the complete absence of any context or information on its type.

Results show that accuracy improves consistently with increasing fragment size. This is in a sense to be expected; all approaches considered in this paper leverage differences between the byte value distribution of random data (which is uniform) and that of compressed data. Perfectly estimating the byte value distribution of a short data stream is generally not possible. As sequences get shorter, the probability that the estimated distribution may not reflect the typical distribution for their content type increases. However, as the size of the sample increases, the estimated empirical distribution approaches the underlying data distribution, allowing us to capture any deviation from the uniform distribution. For modern compression algorithms, these deviations are quite minor, and a 512-byte block gives even accurate tests very little data to work with. However, when enough data is available, it is possible to identify the class of data with high accuracy; our learning-based classifier exceeds 90\% accuracy already for 2048-byte blocks. In general, we recommend against using any one approach as the sole guidance for automated security decisions (e.g. dropping/allowing flows, terminating processes, etc.). However, when integrated as part of a more complete set of features in a larger system, our proposed classifiers can provide an additional robust feature to use in the decision-making process.

Given the discussion above, we suspect an intrinsic bound on the accuracy reachable by any classifier which looks purely at byte value distributions. However, approaches attempting to parse fragments or identify recognizable structures are likely to incur an impractical computational cost. Moreover, it is not apparent that any such structure is preserved for very short fragment sizes.


\section{Related Work}
\label{sec:RelatedWork}

\subsection{Entropy-based encryption detection}

Use of entropy estimation to detect encrypted content is common in ransomware detection. Proposals such as RWGuard~\cite{mehnaz_rwguard}, UNVEIL~\cite{kirda_unveil}, Redemption~\cite{kirda_redemption} and ShieldFS~\cite{continella_shieldfs:_2016} use entropy of written content either directly as a feature, or as part of feature calculation. It should be noted that none of these detectors use entropy as the sole feature for detection. However, evidence from Section~\ref{sec:Background} suggests that they may be better ignoring entropy altogether.
In the realm of digital forensics, entropy estimation has been used to determine the type of unknown disk data fragments. One of the most complete approaches is that of Conti et al.~\cite{conti_automated_2010}. However, the same authors found that such estimates have limited discerning power in distinguishing encrypted and compressed content, and aggregated the two types under a single label.

Entropy estimation has also been applied to the real-time analysis of network traffic. Dorfinger's Master thesis~\cite{dorfinger_real-time_2010} proposes a system for discriminating encrypted and non-encrypted traffic, to ensure that all communications from a target network are encrypted. Similar approaches were also proposed by Mamun et al.~\cite{mamun_entropy_2016} and Malhotra~\cite{malhotra_detection_2007}. Zhang et al. proposed an entropy-based classifier for the identification of botnet traffic~\cite{zhang_detecting_2013}. All these approaches also suffer from the limitations of using high entropy as a fingerprint of encryption.
Wang et al.~\cite{wang_using_2011} report positive results in using an SVM classifier to discriminate between various data types using entropy estimates. Their application scenario is different from ours, as they consider both low-entropy (non-compressed) and high-entropy (compressed or encrypted) formats. We only consider high-entropy formats, which are difficult to distinguish using entropy alone.

Finally, MovieStealer~\cite{wang_steal_2013} aims at identifying encrypted and decrypted-but-compressed media buffers in order to break DRM. It uses an entropy test to single out encrypted and compressed buffers from other data, and the $\chi^2$-test to distinguish them. It requires 800KB of data to reliably identify random data, which is far beyond the fragment size in the scenarios that we consider.

\subsection{Non-entropy-based approaches}

HEDGE, by Casino et al.~\cite{casino_hedge_2019}, evaluates a combination of $\chi^2$-test and a subset of NIST SP800-22~\cite{rukhin_statistical_2010} to discriminate encrypted and compressed traffic. They use a dataset which is significantly smaller than ours, and do not discuss learning-based approaches. A limitation of this class of approaches is the fairly low accuracy, especially for small block sizes (ref. Section~\ref{sec:Evaluation}). Also, this and other similar approaches based on statistical randomness tests (e.g.,~\cite{lipmaa_data_2017,choudhury_empirical_2020}) cannot distinguish between different types of compressed archives. Mbol et al.~\cite{foresti_efficient_2016} investigate the use of the Kullback-Leibler divergence (relative entropy) to differentiate encrypted files from JPEG images. Their analysis does not investigate other formats, and assumes the availability of blocks of significant size (128 to 512KB) from the beginning of each file. Especially in forensic and networking applications, uninterrupted blocks of such size are difficult to obtain.

While the application of neural networks to the problem at hand is fairly new, there exist some preliminary work. Ameeno et al.~\cite{ameeno_using_2019} show promising preliminary results, however the analysis is limited in scope: it only attempts to distinguish zip archives from rc4-encrypted data, and considers whole files (not fragments). Hahn et al.~\cite{hahn_detecting_2018} perform an exploratory analysis of machine learning models. Their dataset is order of magnitudes smaller than ours, and they lack a comparative analysis of statistical approaches.




\section{Conclusions}
\label{sec:Conclusions}

Discriminating encrypted from non-encrypted content is important for a variety of security applications, and oftentimes tackled via entropy estimation. We comprehensively highlighted the limits of this technique and reviewed the effectiveness of the leading alternative approaches: $\chi^2$-test, NIST SP800-22 test suite, and HEDGE. In addition, we proposed \textsc{EnCoD}, a novel neural network classifier of our own design. In order to ensure generality of results, we created a dataset of 200M fragments covering 5 different sizes and 8 data formats.

Results show that previous state-of-the-art methods have blind spots which result in low accuracy for certain fragment sizes/data types. However, our neural network-based approach appears promising. Besides being able to discriminate between compressed formats (which traditional statistical tests cannot), it exceeds 90\% accuracy already on fragments of only 2KB. This suggests that systems incorporating encrypted content detection (e.g., ransomware detectors) would be better served by learning-based, rather than hand-crafted statistical approaches. This finding also suggests that learning may have useful applications to other problems in content type inference. Overall, we believe this work is an important step forward towards reliable encryption detection.


\bibliographystyle{splncs04}
\bibliography{bibliography}
\appendix
{\noindent \bf \Large Appendix}
\section{Entropy Analysis Results}
\label{sec:EntropyAnalysisResults}

\noindent Full results for the entropy analysis discussed in Section~\ref{sec:Challenges}:

\centering
\vspace{0.1in}
  \small
  \begin{tabular}{|l|c|c|c|c|c|}
    \hline
    \multicolumn{6}{|c|}{\textbf{Chunk size: 512B}} \\
    \hline
    \textbf{Format} & \textbf{Min} & \textbf{Q1} & \textbf{Median} & \textbf{Q3} & \textbf{Max}\\
    \hline
    enc & 7.427 & 7.569 & 7.591 & 7.613 & 7.709 \\
    \hline
    zip & 7.163 & 7.560 & 7.584 & 7.607 & 7.695\\
    \hline
    gzip & 7.154 & 7.560 & 7.585 & 7.607 & 7.703\\
    \hline
    rar & 7.381 & 7.563 & 7.587 & 7.610 & 7.692\\
    \hline
    jpeg & 3.820 & 7.512 & 7.548 & 7.576 & 7.676\\
    \hline
    mp3 & 0.000 & 7.451 & 7.527 & 7.565 & 7.680\\
    \hline
    png & 0.000 & 1.070 & 2.605 & 4.549 & 7.572 \\
    \hline
    pdf & 0.000 & 7.453 & 7.534 & 7.574 & 7.676\\
    \hline
    \multicolumn{6}{|c|}{\textbf{Chunk size: 2048B}} \\
    \hline
    \textbf{Format} & \textbf{Min} & \textbf{Q1} & \textbf{Median} & \textbf{Q3} & \textbf{Max}\\
    \hline
    enc & 7.873 & 7.903 & 7.908 & 7.914 & 7.938\\
    \hline
    zip & 7.816 & 7.898 & 7.904 & 7.910 & 7.935\\
    \hline
    gzip & 7.847 & 7.898 & 7.904 & 7.910 & 7.933\\
    \hline
    rar & 7.795 & 7.900 & 7.905 & 7.911 & 7.933\\
    \hline
    jpeg & 5.123 & 7.856 & 7.873 & 7.884 & 7.917\\
    \hline
    mp3 & 0.379 & 7.703 & 7.838 & 7.871 & 7.916\\
    \hline
    png & 0.000 & 1.312 & 2.815 & 4.752 & 7.808\\
    \hline
    pdf & 0.000 & 7.820 & 7.875 & 7.893 & 7.930\\
    \hline
    \multicolumn{6}{|c|}{\textbf{Chunk size: 8192B}} \\
    \hline
    \textbf{Format} & \textbf{Min} & \textbf{Q1} & \textbf{Median} & \textbf{Q3} & \textbf{Max}\\
    \hline
    enc & 7.969 & 7.976 & 7.978 & 7.979 & 7.984 \\
    \hline
    zip & 7.955 & 7.973 & 7.975 & 7.976 & 7.983\\
    \hline
    gzip & 7.955 & 7.973 & 7.975 & 7.976 & 7.983\\
    \hline
    rar & 7.960 & 7.974 & 7.976 & 7.977 & 7.983\\
    \hline
    jpeg & 5.646 & 7.930 & 7.945 & 7.952 & 7.967\\
    \hline
    mp3 & 0.497 & 7.789 & 7.918 & 7.942 & 7.971\\
    \hline
    png & 0.014 & 1.451 & 2.963 & 4.852 & 7.914\\
    \hline
    pdf & 0.010 & 7.903 & 7.953 & 7.968 & 7.981\\
    \hline
  \end{tabular}



\end{document}